\documentclass[smallcondensed]{svjour3}    
\smartqed  
\usepackage{graphicx}
\begin{document}

\title{Trojan capture by terrestrial planets}

\titlerunning{Terrestrial Trojans} 

\author{Schwarz, R.         \and
        Dvorak, R.}


\institute{R. Schwarz \at
              T\"urkenschanzstrasse 17, 1180 Vienna \\
              Tel.: +43-1-427751841\\
              Fax: +43-1-42779518\\
              \email{schwarz@astro.univie.ac.at\\}
\\ 
            R. Dvorak \at
              T\"urkenschanzstrasse 17, 1180 Vienna \\
              Tel.: +43-1-427751841\\
              Fax: +43-1-42779518\\
              \email{dvorak@astro.univie.ac.at}
} 
    
\maketitle

\begin{abstract}
The paper is devoted to investigate the capture of asteroids by Venus, 
Earth and Mars into the 1:1 mean motion resonance especially into Trojan orbits. 
Current theoretical studies predict that Trojan 
asteroids are a frequent by-product of the planet formation. This is not only 
the case for the outer giant planets, but also for the terrestrial planets in
the inner Solar System.  
By using numerical integrations, we investigated the capture efficiency and
the stability of the captured objects. We found out that the capture
efficiency is larger for the planets in the inner Solar System compared to the 
outer ones, but most of the captured Trojan asteroids are not long term stable. 
This temporary captures caused by chaotic behaviour of the objects were investigated 
without any dissipative forces. They show an interesting dynamical behaviour of
mixing like jumping from one Lagrange point to the other one.

\keywords{Trojan capture \and terrestrial planets \and
Near Earth asteroids \and 1:1 mean motion resonance}

\end{abstract}

\section{Introduction}
\label{intro}

In February 1906, Max Wolf~\cite{wolf} discovered the first Trojan asteroid 
named 588 Achilles, after the hero of the Trojan war.  
Almost a century later, in 1990, the first Martian Trojan (5261 Eureka) 
was observed by Bowell et al. \cite{bowell}. In 2001 the first 
Neptune Trojan was discovered by the Lowell Observatory Deep Ecliptic Survey 
team (see Wasserman et al.\cite{wasser}).
Theoretical studies predict that Trojan asteroids are a 
frequent byproduct of planet formation and evolution. A simulation of the chaotic 
capture of Jupiter Trojans (Morbidelli et al. \cite{morbi}) showed that about 
3.4 Earth masses of planetesimals could be captured. Further investigations on the 
formation of (Trojan) planets in the 1:1 mean motion resonance (MMR), as a result 
of an interaction with the protoplanetary disk were done by Laughlin \& Chambers 
\cite{laugh}, Beaug\'e et al. \cite{beauge}, Thommes \cite{thommes}, 
Cresswell \& Nelson \cite{cresswell} and Lyra et al. \cite{lyra}. 
Another important topic concerning the evolution, is the stability of such Trojan 
objects. Several authors have done dynamical studies for Trojan asteroids in the 
solar system (=SS) (e.g. Robutel~\cite{robutel}, Marzari \& Scholl~\cite{marzari}, 
Dvorak et al. \cite{dvorak07}, Freistetter \cite{freistetter}). 
Some of these studies were dedicated to inclined Trojans 
(e.g. Schwarz et al.\cite{schwarz04} and Dvorak \& Schwarz\cite{dvorak05}).
An interesting topic is also the possibility of Trojan planets in extrasolar 
planetary systems like it is discussed e.g. in the dynamical investigations of 
Nauenberg~\cite{nauen}, \'Erdi et al.~\cite{erdi05}, 
Dvorak et al.~\cite{dvorak04}, and Schwarz et al.~\cite{schwarz07}.
From the observational point of view we know approximately 5000 Jupiter Trojans,
7 Neptune Trojans in the outer Solar System, respectively 5 Martian Trojans 
(+1 candidate) and 2 co-orbital objects (horseshoe orbits) close to the Earth in the 
inner Solar System (shown in Table~\ref{trojans}). 
Recently also an asteroid $2010~TK_7$ has been observed,
which librates around the leading Lagrange point $L_4$ of the Earth 
(Connors et al.~\cite{connors}). Calculations showed that the asteroid jump between $L_4$ and $L_5$; 
a so called "Jumping Trojan" which was found in the work of Tsiganis, Dvorak \& Pilat-Lohinger~\cite{tsiganis}.
 
\begin{table}
\centering
 \caption{All observed Trojan asteroids in the inner Solar system.* depicts an
object which is only a candidate.}
  \begin{tabular}{lllll}
  \hline
 Name       & a [AU]   & e  & i [deg] & motion Type\\
  \hline
&&{\bf Martian objects}&&\\
\hline
5261 Eureka & 1.523 & 0.065 & 20.3 & $L_5$ \\
1998 VF31   & 1.524 & 0.100 & 31.3 & $L_5$ \\
1999 UJ7    & 1.524 & 0.390 & 16.8 & $L_4$ \\
2007 NS2    & 1.524 & 0.054 & 18.6 & $L_5$ \\ 
1998 SD4    & 1.523 & 0.314 &  5.6 & horseshoe \\
1998 QH56*  & 1.551 & 0.309 & 32.2 & $L_5$\\
  \hline
&&{\bf Earth objects}&&\\
  \hline
2010 $TK_7$  & 1.000 & 0.191 & 20.880 & $L_4$\\
2002 AA29    & 1.000 & 0.012 & 10.739 & horseshoe\\
3753 Cruithne& 0.998 & 0.515 & 19.811 & horseshoe\\
  \hline
\end{tabular}
\label{trojans}
\end{table}

For the origin of NEAs (e.g. Greenberg \& Nolan~\cite{greenberg1} and 
\cite{greenberg2}) it has been suggested that collisions in the main-belt 
continuously produce new asteroids by fragmentation of larger bodies. 
These fragments can be injected into the $\nu_6$ and 3:1 MMR,
which cause a change of their eccentricities and bring them into orbits 
intersecting the orbit of Mars (Marscrosser) and/or Earth (Earthcrossers, 
e.g. Milani et al.~\cite{milani}).
Farinella et al.,~\cite{farinella} showed that bodies in the $\nu_6$, 3:1 or 
5:2 MMR can have many collisions. Additional studies of the NEAs were done 
by Dvorak \& Pilat-Lohinger~\cite{dvorak99} and Bottke et al.~\cite{bottke}
who investigated their orbital distribution.

The most complete studies concerning the stability of Trojans of the planets in the 
SS have been undertaken by Mikkola \& Innanen~\cite{mikkola}  and
for the inner SS by Tabachnik \& Evans~\cite{taba}, Brasser \&
Letho~\cite{brasser} and for Mars by Scholl \& Marzari~\cite{scholl}. 

The work of Mikkola \& Innanen~\cite{mikkola} (a continuation of a
series of papers by the same authors and also Zhang \& Innanen~\cite{zhang}) dates
back to the 1990 -- with less computer facilities than we have today -- but
still provides important first results. They treated the problem of the
stability of Trojans in the outer and inner SS; for the terrestrial planets
they found that Mercury lose its Trojans within 1~Myrs, whereas Venus, Earth and Mars
Trojans could survive up to 10~Myrs.   

Tabachnik \& Evans~\cite{taba} confirmed partly the earlier results by Mikkola \&
Innanen~\cite{mikkola} for Mercury, but they found
orbits to survive up to 100 Myrs. These stable orbits were not tadpole orbits but all
of them turned out to be on horseshoe orbits, in fact no long-lived Mercurian
Trojans were found. Mercury, having by far the largest eccentricity shows the lowest
probability of hosting Trojans.

The investigations of the Venus Trojan showed that none of them could
survive for inclinations $i > 16^{\circ}$. Like for Venus also for
Earth Trojans  low inclined captured asteroids survived, but for the Earth a
second possible stability zone opens between $16^{\circ} < i <
24^{\circ}$. The second window may disappear for longer integrations; for
both planets some orbits may survive up to several Gyrs.
For Mars low inclined asteroids do not survive, but bodies with inclinations
in the range of $14^{\circ} < i < 40^{\circ}$ move in orbits
which could be stable for up to Gyrs. The instabilies are caused by secular
resonances (=SR) seem to be responsable for ejections. 

In their work Brasser \& Letho~\cite{brasser} carefully checked the role of SR for high 
inclined Trojan asteroids. They found that the Kozai resonance
with Jupiter  (acting for $i > 16^{\circ}$) leads to instability; in addition
many SR between the terrestrial bodies have a destabilizing
effect. Nevertheless they could confirm that the Trojan regions of Venus and Earth
are long-term stable for low inclinations. The trapping into a SR is in the
order of only 0.1~Myrs for Mercury but for the other planets about 1~Myrs.

Scholl \& Marzari~\cite{scholl} studied the Mars Trojans with the aid of the frequency
analysis. Due to the SR the low inclined region is depleted
within Myrs, and stable regions exist for inclinations $15^{\circ} < i <
30^{\circ}$, where also the discovered Mars Trojans are moving. 

According to these results we decided to investigate the capture into 1:1 MMR
in the inner SS except Mercury with special emphasis temporary captures into
Trojan motion.

\section{Model and methods of investigation}

The aim of this work is different from former studies which wanted to
establish stable zones of Trojans of the planets. We investigated the capture
of small bodies into the 1:1 MMR with the planets Venus, Earth and
Mars. Therefore we did N-body simulations in a simplified dynamical model,
which consists of Venus, Earth, Mars, Jupiter 
and Saturn up to 10~Myrs. Mercury was excluded from our computation because this
planet does not seem to be able to host long lived
Trojans (Tabachnik \& Evans~\cite{taba}) and the inclusion of Mercury would slow
down the speed of the computation significantly; this  means that the needed
computer time would be increased significantly.
We used the Lie-method with an automatic step-size control to solve the 
equations of motion (Hanslmeier \& Dvorak~\cite{hanslmeier}, Delva~\cite{delva}, 
Lichtenegger~\cite{licht}).
The massless asteroids where placed in 3 different regions, which 
represent almost the whole region of the NEAs:

\begin{itemize}
\item Region A: $0.72<a<0.98$ ($\sim$ Atens)
\item Region B: $1.02<a<1.50$ ($\sim$ Apollos)
\item Region C: $1.54<a<2.20$ ($\sim$ Amors, and the main belt group Hungarias)
\end{itemize}

The sorting of the NEAs into different groups was introduced by Shoemaker
et al.~\cite{shoemaker}.
Milani et al.~\cite{milani} classified the asteroids according to
their collision probabilities and close encounters, whereas
Freistetter~\cite{freistetter} used methods from Fuzzy Logic.

In our studies we distributed 100 asteroids uniformly in the three different 
regions defined above. The initial eccentricities were set close to the
planet's eccentricity and higher (e=0.1 and e=0.2). The choice of the
  other initial conditions was a random one: because we did
the integrations starting with e=0.1 (close to Mars e=0.098) we used for the
other elements initially (all the angles) the ones of Mars. We are aware 
that this choice is more or less artificial but during the integration 
of $10^7$ years this choice does not seem to make big differences 
statistically. In Fig.~\ref{dist} we show that for the initial conditions 
(e=0.1 or e=0.2) the distribution of eccentricities is
very similar for both at the end of integration (with a peak at e=0.25)
and the bodies achieve large eccentricities. We started from almost plane
orbits up to large inclinations; thus we have chosen the following initial
values: i=$1^{\circ}$, $4^{\circ}$, $9^{\circ}$, $16^{\circ}$, $25^{\circ}$,
$30^{\circ}$ and $36^{\circ}$). Because we did not find any captures for
  $i=36^{\circ}$ we made an additional run for $i=30^{\circ}$. All in all 
(test computations included) some ten-thousand orbits of the asteroids were computed.
The analysis of the data showed that we have to distinguish between different 
types of captures (see Figs.~\ref{type} and \ref{moon}):

\begin{enumerate}
\item Satellite orbits 
\item Tadpole orbits ($L_4$ or $L_5$)
\item Jumping Trojans\footnote{The asteroids jump from $L_4$ to $L_5$ or vice versa, 
Tsiganis et al.~\cite{tsiganis}}
\item Horseshoe orbits
\end{enumerate}

During our studies we found several single and multiple captures (can happen
for one object). In case of multiple captures we observed a sort of mixing,
between the different types of captures:
jumping from one Lagrange point to
the other, developing from a tadpole to a horseshoe orbit and vice-versa,
transiting from horseshoe orbits to satellites of the planet and vice versa.
The classification was done by checking the libration width $\sigma$ which
is defined as the difference between the mean longitude of the asteroid and the 
planet (Venus, Earth or Mars) ($\lambda -\lambda_P$). $\lambda$, $\lambda_P$ are 
given by $\lambda=\varpi+M$, $\lambda_P=\varpi_P+M_P$ were $\varpi$, $\varpi_P$ 
are the longitudes of the asteroid and of the planet and $M$, $M_P$ are the mean 
anomaly of the asteroid respectively of the planet.

\section{Results}

\begin{figure}[h]
\centerline{
\includegraphics[width=8.5cm,angle=270]{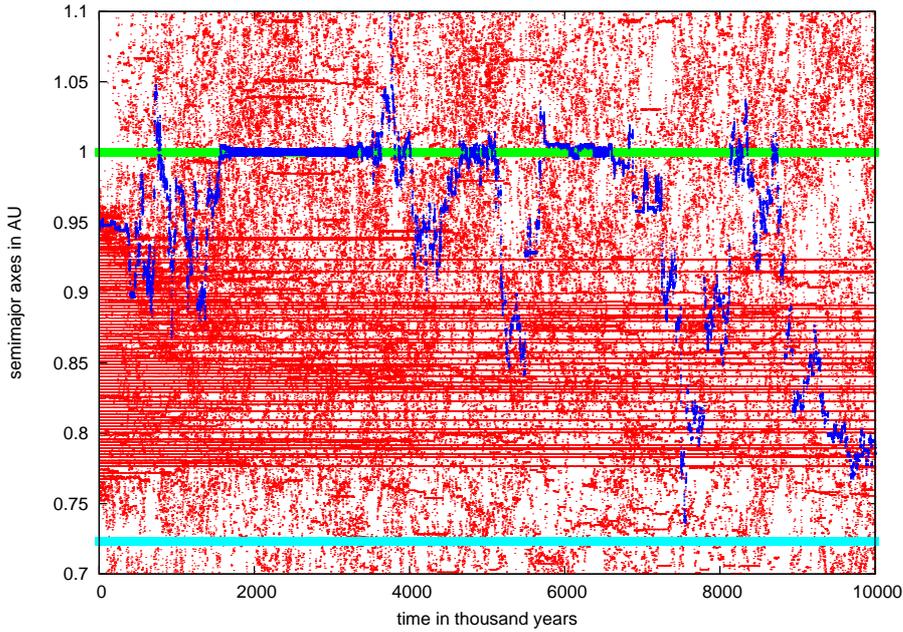}}
\caption{Example of one of the runs for the region A where 100 asteroids were
  located between Venus and Earth. We show the dynamical evolution of the
semimjaor axes of  celestial bodies for 10~Myrs. Venus' and Earth's semi-major
axes are just straight lines for a=0.72~AU and a=1~AU}
\label{sample}
\end{figure}

\begin{figure}[h]
\centerline{
\includegraphics[width=8.5cm,angle=270]{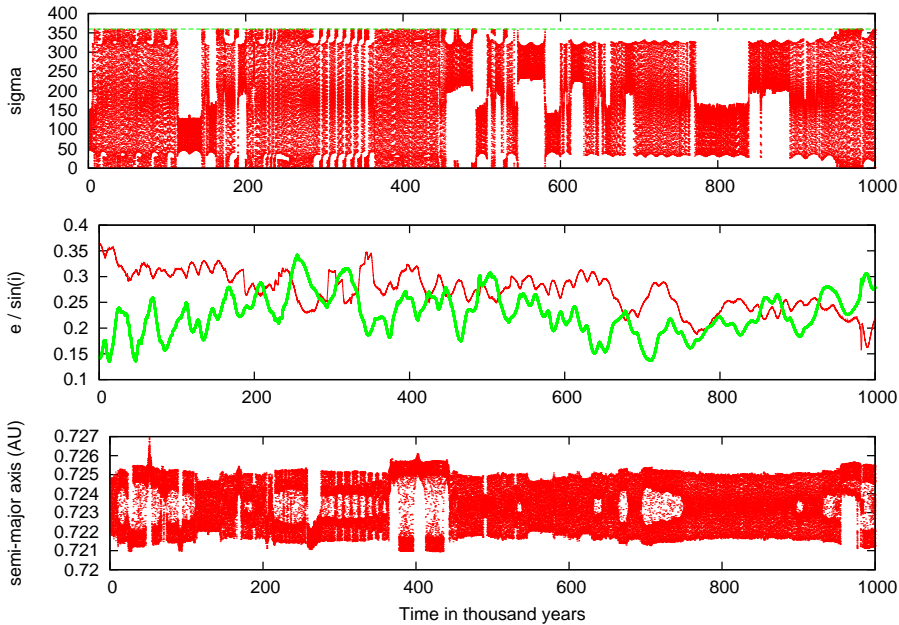}}
\caption{Dynamical evolution of an asteroid captured by Venus: the libration
  angle  $\sigma$ (upper panel), the eccentricity (green line), starting as lower
  line ~0.15) and the inclination given in values of sin(i) (red line) (middle panel)
and the semi-major axis for 1~Myrs (lower panel); for a detailed description see the text.}
\label{type}
\end{figure}

\begin{figure}[h]
\centerline{
\includegraphics[width=5.5cm,angle=270]{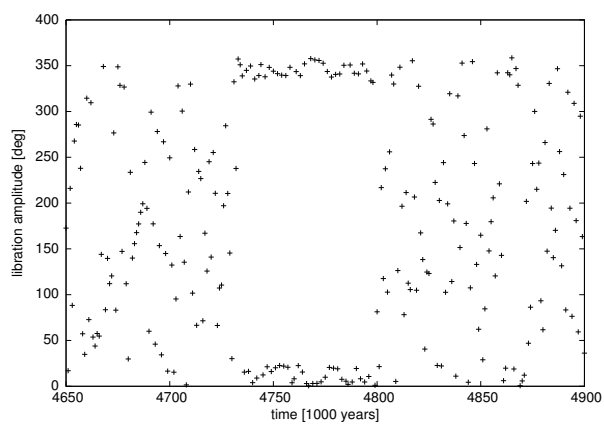}}
\caption{Dynamical evolution of the libration angle  $\sigma$ of an
  asteroid captured by Venus into a satellite
  configuration.}
\label{moon}
\end{figure}

\begin{figure}[h]
\centerline{
\includegraphics[width=5.5cm,angle=270]{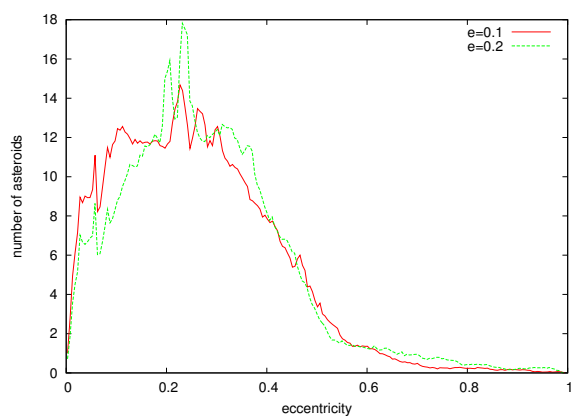}}
\caption{Distribution of the maximum eccentricities of the  fictitious
    asteroids with initial values $e=0.1$ (red line) and $e=0.2$ (green line)
    during their dynamical evolution}
\label{dist}
\end{figure}

In Fig. 1 we depict as one example the dynamical evolution of 100 fictitious
bodies in region A with initial condition e=0.2 and i=0.1. There are two
candidates visible for captures by the Earth, one very short one ($T \sim
9.6~Myrs$) and one very long one ($1.8~Myrs < T <  4~Myrs$), where all
different kinds of captures may occured. 
 We will discuss in more detail the verified captures which are 
presented Fig.~\ref{type} in the next paragraph. The capture candidates
by Venus are visible for two short intervals of time (1.8 Myrs  $< T <$
1.9 Myrs and 4.6 Myrs $< T <$ 5 Myrs) by two different asteroids. There are some
other interesting features like long straight lines for the whole time 
interval-- just
stable orbits in the region A -- and other captures into MMR where the
semi-major axis jumping and stay close for quite a long time; one
is well visible for $a_{ini} =0.925$~AU for $T  \sim$ 6.3~Myrs.  Other such captures
into high order MMR are visible for orbits outside the orbit of the Earth
(e.g. a=1.05~AU for 1.8~Gyrs $< T <$ 4~Myrs. ) and inside the orbit of
Venus and Earth (e.g. a=0.5~AU for 8.0~Myrs $ < T <$  8.5~Myrs).

Fig.~\ref{type} shows the multiple events of the dynamical evolution of
one asteroid captured by Venus and present the different cases of captures (mixed capture).
In the upper panel of Fig.~\ref{type} the libration angle is depicted for 1~Myrs.
The body is captured approximately after 0.13~Myrs, leaves then 
the region around $L_4$ for about 0.05~Myrs but is captured very soon after again for several
thousand years with a slightly larger libration. Close to 0.2~Myrs the
fictitious asteroid is captured around $L_5$. After about 0.42~Myrs the
asteroid jumps from one libration point to the other one. Around 0.8~Myrs the jumps happen in the other
direction, namely from $L_5$ to $L_4$ which ends around 0.9~Myrs with a
horseshoe orbit. Transient capture for being a satellite are only for short
time of several thousand years (e.g. around 0.19~Myrs). The inclination
(middle panel) varies between $5^{\circ} < i < 20^{\circ}$ whereas the eccentricity is
relatively large and changes between $0.15 <  e < 0.35$. Evidently there
is no correlation between these two orbital elements, which is
quite contrary to planetary orbits where the angular momentum forces such a
correlation; note that the fictitious asteroid is regarded as having no mass. The orbital elements 
(e, i and a) do not show whether the body is captured in a Trojan configuration. The discussion of 
the libration  angle shows for the different types of captures the following characteristics: for the 
satellite type $\sigma$ is around $0^{\circ}$, for $L_4$ $\sigma \approx 60^{\circ}$, for $L_5$ $\sigma 
\approx 270^{\circ}$. For horseshoe type the amplitude of the libration width is larger 
$\Delta \sigma \approx 300^{\circ}$, resulting in an orbit where the asteroid passes $L_4$, $L_3$ and 
$L_5$ or vise versa, as well as the jumping Trojans. 

The captures from the other planets show quite a similar dynamical behaviour.
But we must say that a capture by a terrestrial planet is not a frequent phenomenon,
however, our results have shown that only 35 captures of all captures (total
captures are 77) in the inner SS are not mixed.
Details of the different planets are shown in Table~\ref{cap}, where we can see that 
approximately one third of all captures are not mixed for the planets Venus and 
Earth, whereas for Mars the half of all captures are not mixed.

We conclude that these captured objects are very chaotic: it is not possible
to reproduce a capture starting from the time just before the capture happens.
 Integrating 100 fictitious asteroids -- as we have done it -- results in many
 changes of the stepsize because one of the
bodies may be close to a planet choosing a small stepsize\footnote{we use an automatic
stepsize control and it is the same for all involved celestial bodies}.
This leads to different accumulation of errors which propagate quite fast in a
chaotic domain for different integrations when one repeats it from a
  later time; slightly different initial conditions. It
means that the 'capture window' is quite small, but still large enough to allow
many of them.

\begin{table}
\centering
\caption{Comparison of the total number of captures and the mixed captures in 
the inner Solar system.}
\begin{tabular}{lll}
\hline
Planet & total   & not mixed\\
       & capture & capture  \\
\hline
Venus   & 21 &  10  \\
Earth   & 44 &  18  \\
Mars    & 12 &   7  \\
\hline
All     & 77 &  35  \\
\hline
\end{tabular}
\label{cap}
\end{table}

\begin{table}
\centering
 \caption{Asteroid captures from region A for Venus (V), Earth (E) and Mars (M); the
   initial eccentricity of the fictitious body is given in parenthesis.}
  \begin{tabular}{lllllll}
  \hline
 incl(deg) & V(0.1)   & V(0.2) & E(0.1) & E(0.2) & M(0.1) & M(0.2) \\
\hline
1 & 1 & 2 & 1 & 2 &- & -\\
4 & 2 & 1 & 2 & 1 & - & -\\
9 & 2 & - & 8 & - &1 & -\\
16 & 3 & - & 4 & 2 & 1 & -\\
25 & 3 & 2 & 1 & 5 & - & -\\
30 & 2 & 1 & 3 & 3 & 2 & - \\
36 & - & - & - & - & - & -\\
\hline
\end{tabular}
\label{tabA}
\end{table} 

\begin{table}
\centering
 \caption{Asteroid captures from region B for Venus (V), Earth (E) and Mars (M); the
   initial eccentricity of the fictitious body is given in parenthesis}
  \begin{tabular}{lllllll}
  \hline
 incl(deg) & V(0.1)   & V(0.2) & E(0.1) & E(0.2) & M(0.1) & M(0.2) \\
\hline
1 & 1 & - & 2 & - &- & -\\
4 & - & - & 1 & - & 2 & -\\
9 & - & 1  & 1 & - & - & 1\\
16 & - & - & 1 & 1 & - & -\\
25 & - & - & - & - & - & -\\
30 & - & - & 2 & 4 & 2 & 2 \\
36 & - & - & - & - & - & -\\
\hline
\end{tabular}
\label{tabB}
\end{table}

As mentioned before we used for our simulations 4200 particles which 
were distributed over all three regions (region A, B and C). 
Our results have shown that from the 4200 asteroids 77 individual asteroids were captured into 
co-orbital motion. 
In total we measured 205 capture events, because of the former mentioned 
mixed capture. 59 capture events were measured close to $L_4$ and 61 close to $L_5$.
Comparable capture events were found in horseshoe motion (52 captures), 24
capture events were found in satellite motion and 9 events happens for 
jumping Trojans. A summary of the captures for the different planets are shown in
Table~\ref{cap}. There we can see that approximately one third of all captures
are not mixed for the planets Venus and Earth, whereas for Mars half of all captures are not mixed.

We found 3 jumping Trojans for the Earth, which is in a good agreement with the 
recent work of Connors et al.~\cite{connors} who observed a $L_4$ Trojan.
Their calculations showed that the orbit of $2010~TK_7$ is at least stable
for 10000 years, which is again in good agreement with our studies. In addition
they computed the orbit of this Trojan asteroid in the past and found a 'jump'
from $L_5$ libration to the present $L_4$ libration at AD 400. In our work we
located 4 asteroids for Venus, 3 for Earth and 2 for Mars to be captured as
jumping Trojans.

\begin{figure}[h]
\centerline{
\includegraphics[width=7.0cm,angle=0]{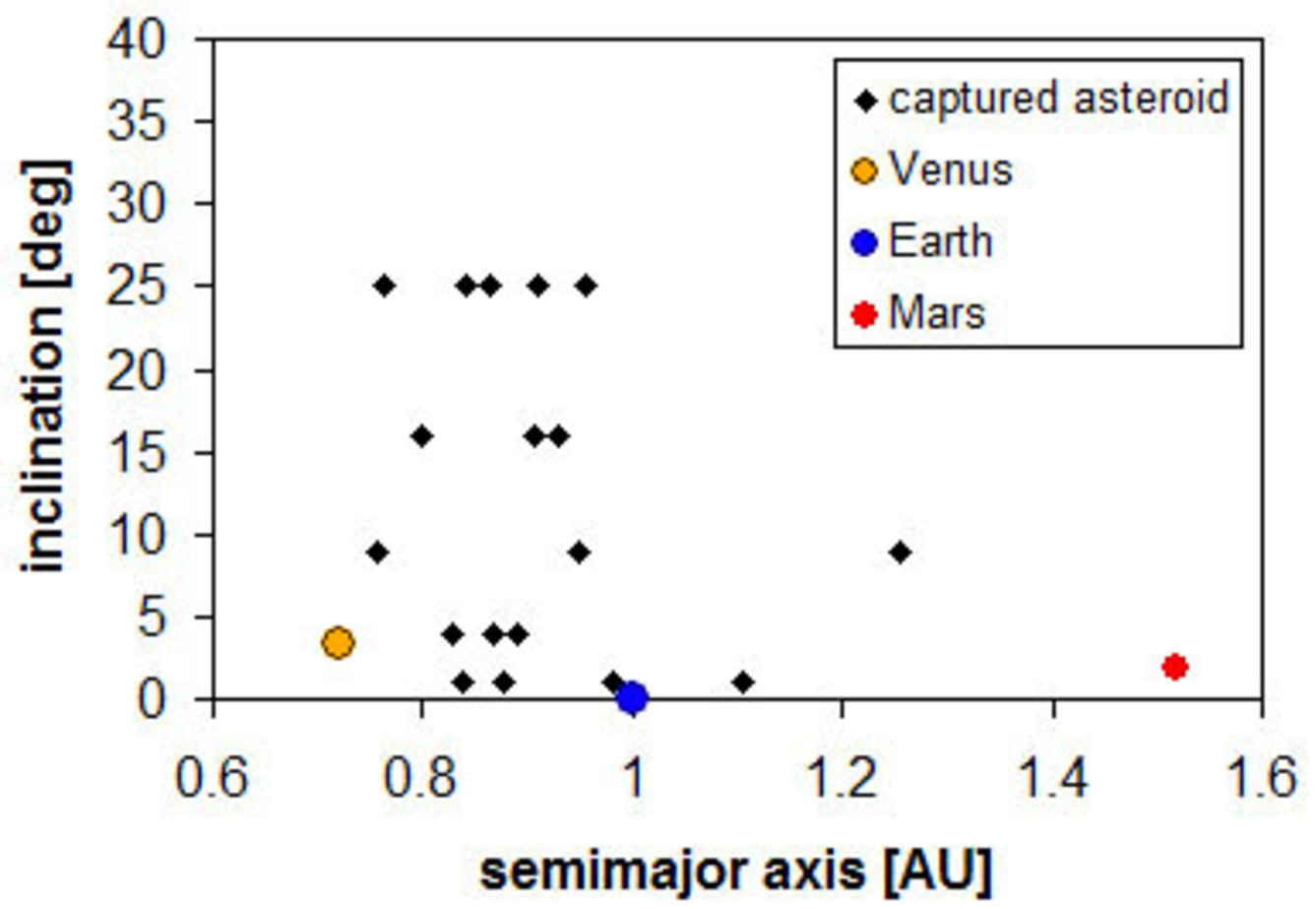}}
\centerline{
\includegraphics[width=7.0cm,angle=0]{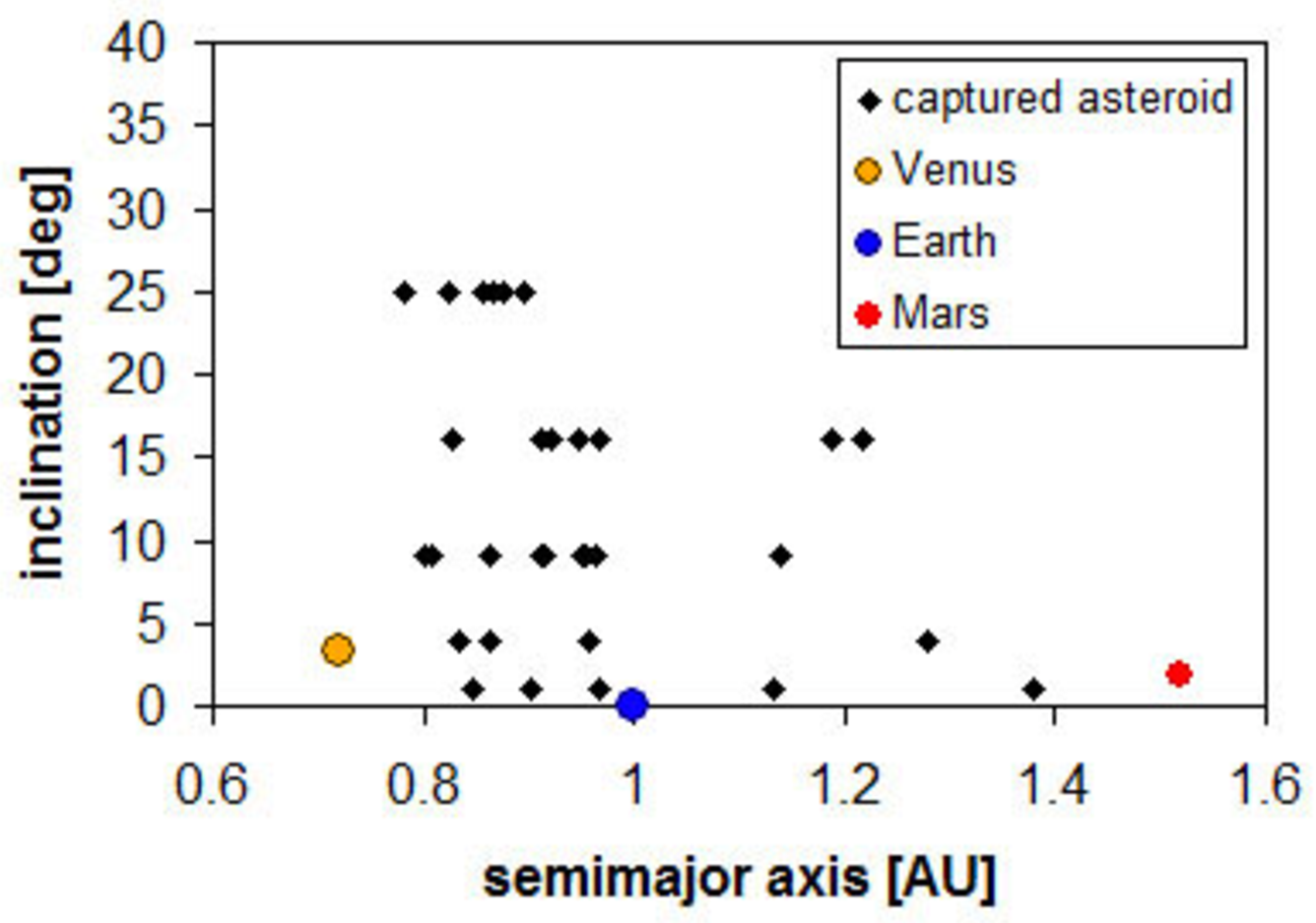}}
\centerline{
\includegraphics[width=7.0cm,angle=0]{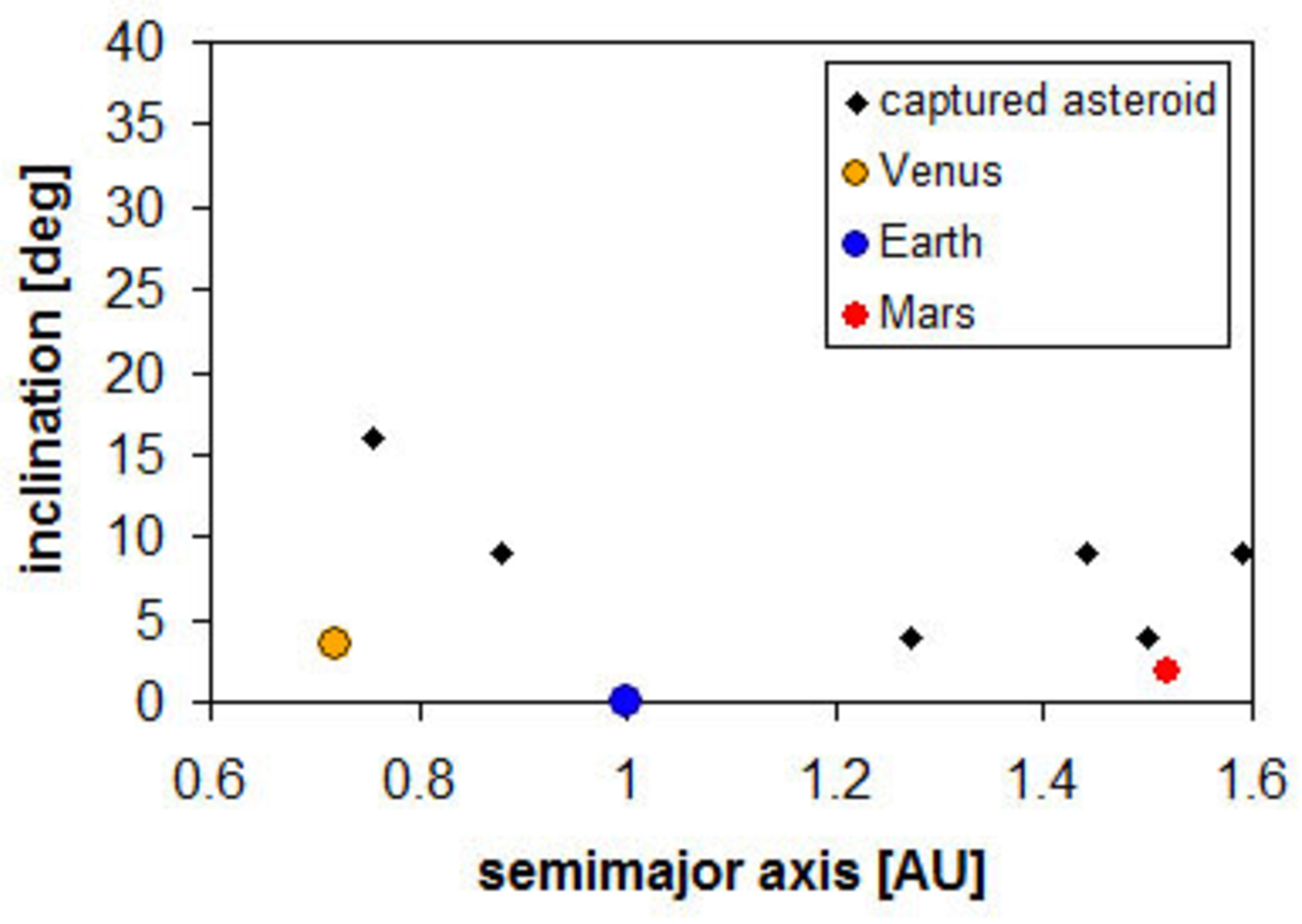}}
\caption{Initial orbital elements semimajor axis versus the inclination of the
captured asteroids by Venus (upper graph), the Earth (middle graph) and by 
Mars (lower graph).}
\label{ini}
\end{figure}

In a next step we wanted to show which 'preferred' initial conditions lead to
a capture into 1:1 MMR. Details are shown in the Tables~\ref{tabA} and \ref{tabB} 
separately for the region A and region B. They are summarized in Fig.~\ref{ini} for the planets
Venus (upper graph), Earth (middle graph) and Mars (lower graph). In case of Venus
most captures come from region A (19) and only 2 captures from region B.
Also Earth captured most asteroids from region A (32) and several from region B
(12). The capture of the Martian co-orbitals can be observed
from all regions: Region A (4), Region B (7) and Region C (1). Further investigations
have to be done to study more in detail the influence from the outer main belt, which will
possibly change the capture probability of Mars. Many captures were found for
large inital inclinations $i_{ini}$ up to $30^{\circ}$, but we did not find
any captures for $36^{\circ}$.

\begin{figure}[h]
\centerline{
\includegraphics[width=7.0cm,angle=0]{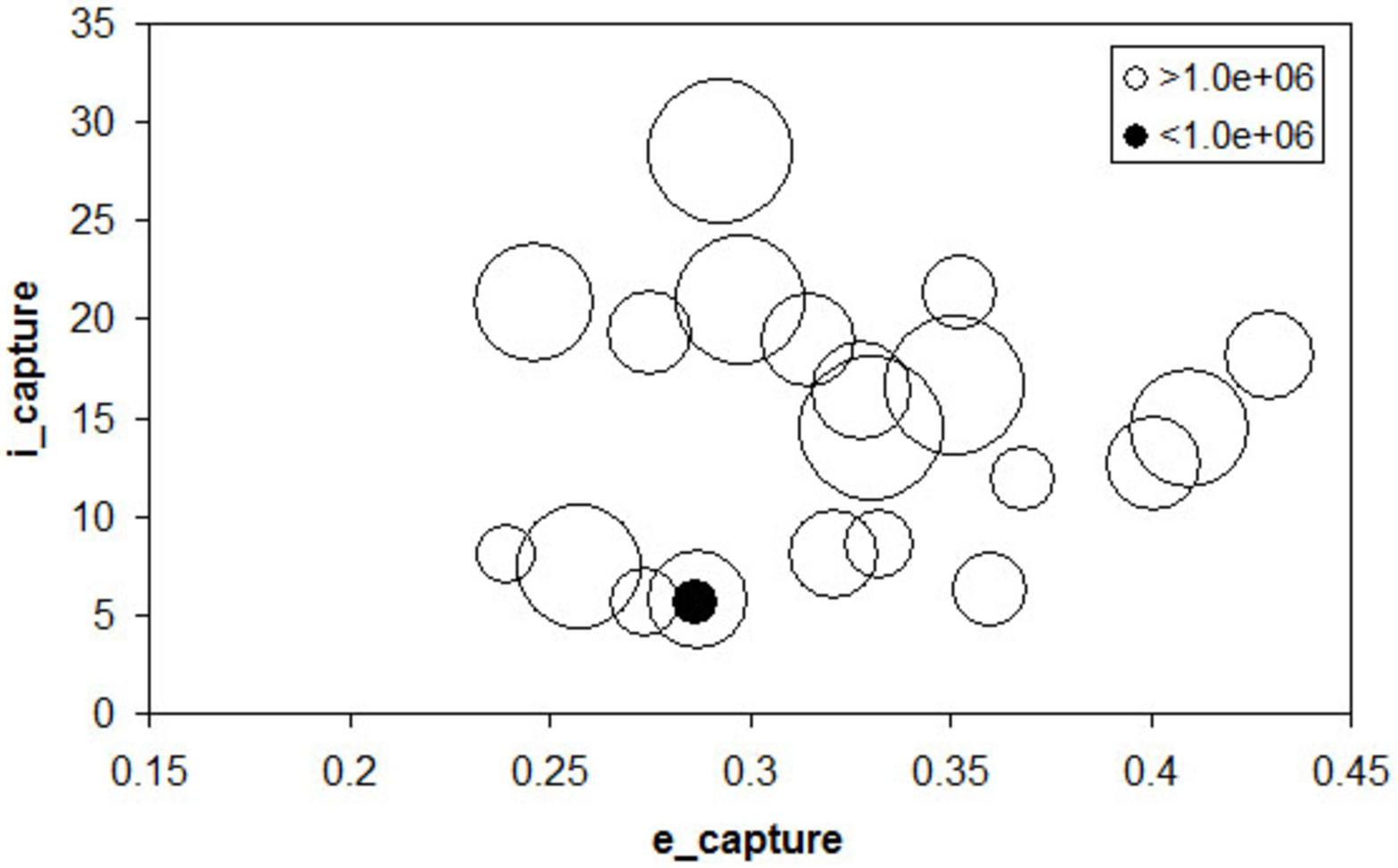}}
\centerline{
\includegraphics[width=7.0cm,angle=0]{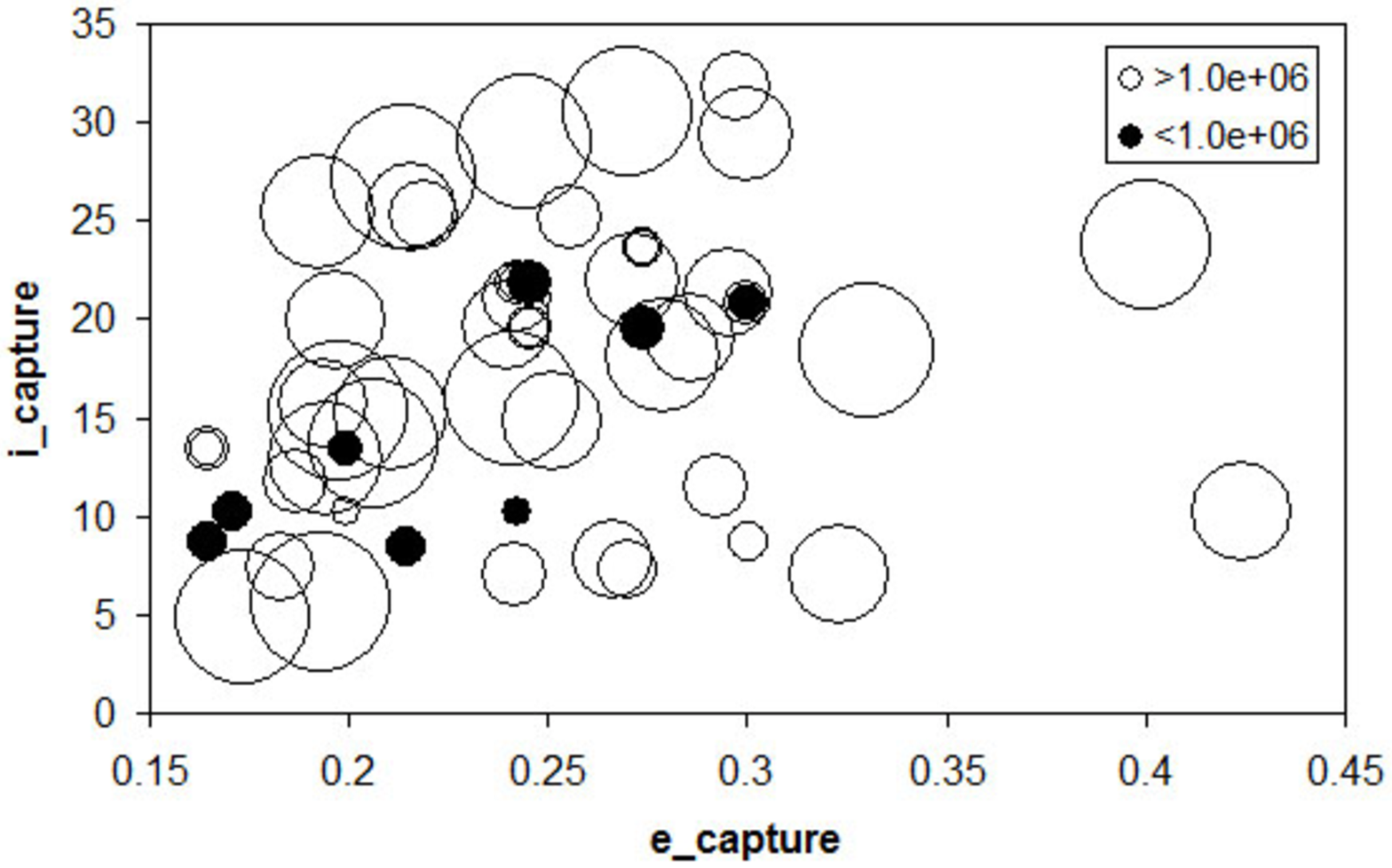}}
\centerline{
\includegraphics[width=7.0cm,angle=0]{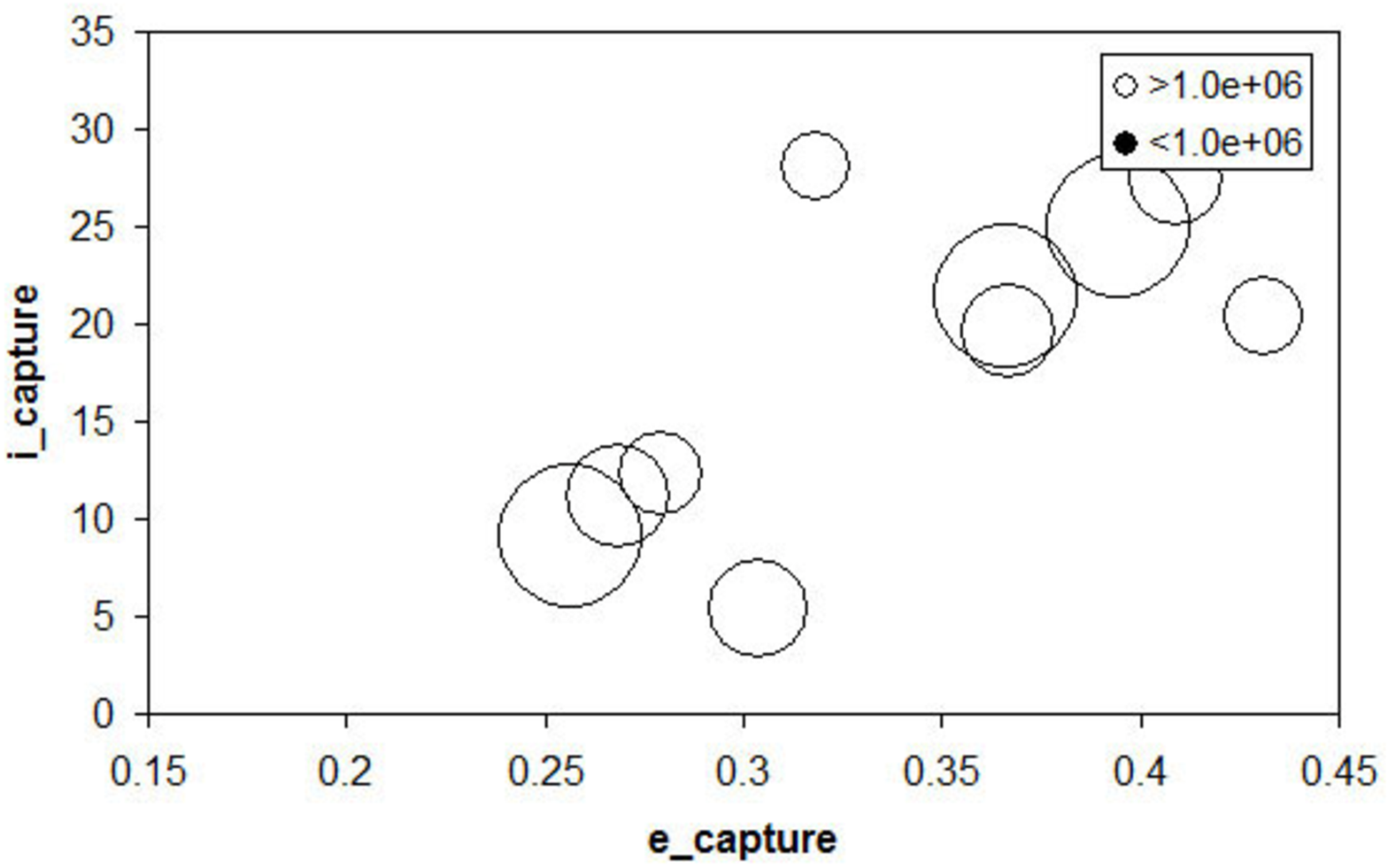}}
\caption{Mean orbital elements (eccentricity and inclination) just before the capture 
into the co-orbital motion  by Venus (upper graph), the Earth (middle graph) and by 
Mars (lower graph). We distinguished 
between a capture time which is smaller and larger than 1~Myrs. The largeness
of the circles shows whether the capture time is large or not.}
\label{capture}
\end{figure}

We also examined the eccentricity and the inclination at the time of a capture.
These results are presented in Fig.~\ref{capture} including the capture time 
of the first capture. We call that first capture, because
in case of mixed capture more captures can happen. We distinguished
between capture times smaller and larger than 1~Myrs.
Our investigations showed that the planetesimals which were started close 
to a planet, will be captured earlier (first capture $< 1~Myrs$).
This statement agrees quite well in case of Venus and 
Earth, where all planetesimals which were captured are within an initial distance
to the planets of $a< 0.2~AU$. However for Mars there is no agreement, because of
the longer time interval up to the first capture Fig.~\ref{capture} (lower panel).
The Earth is presented in Fig.~\ref{capture} middle graph, which shows that most captures
are for asteroids initially close to the planet (presented as black circles in
the graph) also visible in the initial condition diagramm of Fig.~\ref{ini} (middle graph). 
In case of Venus we have only 1 close capture and for Mars we 
have 3 close capture, because the captures are distributed (as mentioned before). 
Furthermore we can conclude that all captures happen in the orbital parameter-space:
$0.15<e_{capture}<0.45$ and $3^{\circ}<i_{capture}<32^{\circ}$.

Another important issue is the stability of the captured asteroids.
We found out that the most stable objects are captured asteroids starting
close to Mars with low initial inclinations ($i=9^{\circ}$).
The maximum and the mean duration of the captured asteroids  into stable co-orbital 
motion for:
\begin{enumerate}
\item Venus was $6.3\cdot 10^5$ years (mean value $1.8\cdot 10^5$ years),
\item Earth $6.5\cdot10^5$ years (mean value $1.6\cdot 10^5$ years)
\item Mars $2.9\cdot 10^6$ years (mean value $3.4\cdot 10^5$ years).
\end{enumerate}
Only three captured objects were still stable after the computations of
the orbits for 10~Myrs,
two close to Earth and one close to Mars. The latter one is the most stable asteroid being
captured into a horseshoe motion after ~7.1 Myrs and stays there ~2.9 Myrs up to the end of the
integration (10 Myrs).
  
\subsection{Comparison of the capture efficiency in the inner and outer Solar system}

The capture efficiency in the outer SS was studied by Morbidelli et
al.~\cite{morbi} in the case of Jupiter Trojans and by 
Nesvorny \& Vokrouhlicky~\cite{nesvorny} for the Neptune Trojans, by using the
Nice model. The Nice model was designed for the investigation of the early evolution of the outer SS. 
In this model, the giant planets are assumed to have 
formed in a compact configuration (5-18~AU). Thus these studies consider also the 
interaction of the planetesimal disk with the planets, which induces a slow migration
of the giant planets.
We want to remark that our model is a simple n-body integration without any gravitational influence
from the asteroid on the planets. Nevertheless, we compared our capture
efficiency with the former mentioned studies. 
The capture efficiency\footnote{The capture efficiency is defined per particle
in the whole initial parameter space like in Nesvorny \& Vokrouhlicky~\cite{nesvorny} and Morbidelli et
al.~\cite{morbi}.} is given in Table~\ref{comparison} ordered by the distance to the Sun.

\begin{table}
\centering
 \caption{Comparison of the capture efficiency from the inner Solar and the 
outer solar system.}
  \begin{tabular}{ll}
  \hline
 Planet    & capture  \\
           & efficiency\\
  \hline
\bf{inner solar system}&\\
\hline
Venus   & $4.3 \cdot 10^{-3}$  \\
Earth   &$1.05 \cdot 10^{-2}$ \\
Mars    & $2.4 \cdot 10^{-3}$ \\
  \hline
\bf{outer solar system}&\\
\hline
Jupiter & $2 \cdot 10^{-5}$ \\
Neptune & $3 \cdot 10^{-4}$ \\
  \hline
\end{tabular}
\label{comparison}
\end{table}

With this overview we can conclude that the capture probability in the inner
SS is higher than in the outer one, especially taking into account that in the
Nice model nongravitational forces were also included in their integration
which possibly enlarged the number of captured Trojans (as presented in Table~\ref{comparison}).

\section{Conclusions}

\begin{table}
\centering
 \caption{The mean, minimum and maximum values of the orbital elements e and i
   for the different groups of NEAs. The data was taken from the Minor Planet Center August
   2011).}
  \begin{tabular}{lll}
  \hline
 Region    & eccentricity & inclination  \\
           &    &  [deg]\\
  \hline
\bf{Atens} & & \\
\hline
mean value & 0.35  & 13.67  \\
minimum    & 0.01  &  0.00  \\
maximum    & 0.90  & 56.10  \\
\hline
\bf{Apollos} & & \\
\hline
mean value & 0.52 &  13.71  \\
minimum    & 0.03 &  0.00  \\
maximum    & 0.97 & 154.50  \\
\hline
\bf{Amors}& & \\
\hline
mean value & 0.42  &  14.27  \\
minimum    & 0.01  &   0.10  \\
maximum    & 0.74  & 131.80  \\
\hline
\end{tabular}
\label{statistic}
\end{table}

Up to now several simulations were done for the outer SS, 
to show the capture of asteroids into co-orbital motion during the time when
the giant planets migrated.
We focused our work on the capture of co-orbital objects in the inner
SS. In our studies we found several captures of asteroids into: tadpole
(including $L_4$, $L_5$ Trojans and jumping Trojans), horseshoe and satellite motion
for all three planets Venus, Earth and Mars. 
The studies up to now undertaken for the terrestrial planets deal primarily
with the stability of samples of fictitious bodies already in a 1:1 MMR with
a planet to establish stable regions around the equilateral equilibrium
points. Normally these computations were stopped after the escape from the
resonance. Our goal was a different one: we started from outside (or inside)
the planets orbit away from the 1:1 MMR and checked for respective captures. 
Because these asteroids are on the edge of the
stability region the 'stability times' (the state when they are in the phase of
being coorbital with the planets) are sometimes very small (see Fig.~\ref{capture}).  
We can conclude from our results that the capture probability in the inner SS 
is higher than in the outer SS (205 capture events), but the asteroids are not
long-term stable (Tabachnik \& Evans~\cite{taba}. 
Nevertheless, the capture window is limited by $0.15 < e_{capture} <
0.45$ and by $3^{\circ} < i_{capture} < 32^{\circ}$, which is valid for
all three planets. 

In fact the mean values for the eccentricities of the three groups of the NEAs
are close to the highest value of e=0.45 in our investigation (compare with
Fig.\ref{capture}). For the inclinations the mean value of the NEAs coincides
with the mean value for the captured Trojans (Tab.~\ref{statistic}).

All the captures we found are transient events; the probability of such a
captured asteroids located initially between the terrestrial planets with
moderate eccentricities and inclinations is surprisingly high. We explain it
that after an integration time of about 1Myrs the eccentricities reach large
values comparable to some of the discovered real Trojans of Mars (like 1999 UJ7).

We detected several multiple capture events showing a mixing of captures. That
means they into different kinds of co-orbital motion. This could be confirmed in the
work of Galiazzo \& Schwarz~\cite{gal14} which was dedicated to investigate this mechanism from the Hungaria family.   
Especially interesting are the quite often observed jumping
objects from one Lagrange point to the other. The first jumping Trojan
was reported by Tsiganis et al.~\cite{tsiganis} for the Jupiter Trojan
$Thersites$ and recently observed by Connors et al.~\cite{connors} for the
Earth Trojan $2010~TK_7$. These very peculiar orbits need to be studied in more 
details not only for SS bodies but also in simplified dynamical models
(e.g. the restricted three-body problem),which is an ongoing work.

\begin{acknowledgements}
R. Schwarz and R. Dvorak wants to acknowledge the support by the Austrian
FWF Project P 18930-N16. Special thanks to V. Eybl and \'A. Bazs\'o who
helped to improve the paper.
\end{acknowledgements}


\end{document}